\documentclass{aa}
\usepackage[english]{babel}
\usepackage{epsfig,color,natbib}
\usepackage[dvips]{rotating}
\usepackage{graphics}
\usepackage{amsmath}

\bibpunct{(}{)}{;}{a}{}{,}

\def\Msun{\hbox{M$_{\odot}$}}               
\def\kms{\hbox{km$\;$s$^{-1}$}}
\def\mic{\hbox{$\mu$m}}                     
\def\Teff{\hbox{$\rm{T}_{\rm eff}$}}            
\def\ad{\hbox{$\theta_d$}}                  

\def\jchemp{J.\ Chem.\ Phys.}

\begin{document}

\title{Theoretical model atmosphere spectra used for the calibration
of infrared instruments}

\author{L.\ Decin \inst{1,2} \thanks{\emph{Postdoctoral Fellow of the Fund for
Scientific Research, Flanders}}
\and K.\ Eriksson \inst{3}}

\offprints{L.\ Decin, e-mail: Leen.Decin@ster.kuleuven.ac.be}

\institute{Department of Physics and Astronomy, Institute for
  Astronomy, K.U.Leuven, Celestijnenlaan 200\,D, B-3001 Leuven, Belgium
  \and Sterrenkundig Instituut Anton Pannekoek, University of
  Amsterdam, Kruislaan 403, NL-1098 Amsterdam, The Netherlands \and
  Department of Astronomy and Space Physics, Uppsala University, Box
  515, SE-751\,20 Uppsala, Sweden}

\date{received date; accepted date}

\abstract{One of the key ingredients in establishing the relation
  between input signal and output flux from a spectrometer is 
  accurate determination of the {\em spectrophotometric
  calibration}. In the case of spectrometers onboard satellites, the
  accuracy of this part of the calibration pedigree is ultimately
  linked to the accuracy of the set of reference spectral energy
  distributions (SEDs) that the spectrophotometric calibration is built
  on.}  {In this paper, we deal with the spectrophotometric
  calibration of \emph{infrared} (IR) spectrometers onboard satellites
  in the 2 to 200\,$\mu$m wavelength range. We aim at comparing the
  different reference SEDs used for the IR spectrophotometric
  calibration. The emphasis is on the reference SEDs of \emph{stellar
  standards} with spectral type later than A0, with special focus on
  the theoretical model atmosphere spectra.} {Using the MARCS model
  atmosphere code, spectral reference SEDs were constructed for a set
  of IR stellar standards (A dwarfs, solar analogs, G9-M0 giants). A
  detailed error analysis was performed to estimate proper
  uncertainties on the predicted flux values.}  {It is shown that the
  uncertainty on the predicted fluxes can be as high as 10\,\%, but in
  case high-resolution observational optical or near-IR data are
  available, and IR excess can be excluded, the uncertainty on
  medium-resolution SEDs can be reduced to 1--2\,\% in the near-IR, to
  $\sim$3\,\% in the mid-IR, and to $\sim$5\,\% in the far-IR. Moreover,
  it is argued that theoretical stellar atmosphere spectra are at the
  moment the best representations for the IR fluxes of cool stellar
  standards.}  {When aiming at a determination of the
  spectrophotometric calibration of IR spectrometers better than
  3\,\%, effort should be put into constructing an appropriate set of
  stellar reference SEDs based on theoretical atmosphere spectra for
  some 15 standard stars with spectral types between A0~V and
  M0~III. }

\keywords{Instrumentation: spectrographs, Techniques: spectroscopic,
  Stars: atmospheres, Stars: late-type, Infrared: stars} 

\titlerunning{Theoretical atmosphere spectra used for IR calibration}
\maketitle


\section{Introduction}\label{introduction}

For (IR) spectrometers, three main steps have to be covered to
establish the relation between input signal, flux, and wavelength:
spatial, spectral, and photometric calibration. In this paper, we 
focus on the reference spectral energy distributions (SEDs) used in
the spectrophotometric (\,=\,spectral+photometric) calibration
process, with main emphasis on the spectral calibration part. 

For (IR) spectrometers onboard satellites, the determination of the
relative spectral response function (RSRF), which characterises the
wavelength-dependent response of a spectrometer, is often a two-step
process: (1) first, the RSRF is determined during
laboratory tests from measuring a cryogenic blackbody calibration
source at differing temperatures; (2) after launch, the RSRF is
refined by comparing observations with reference SEDs of various
celestial calibration sources, preferably at a spectral resolution
comparable to that of the instrument.

Whenever an instrument covers a new spectral window or has a higher
sensitivity than preceding instruments, new reference SEDs have to be
constructed. This was and will be the case for instruments onboard the
ESA-Infrared Space Observatory (ISO), the NASA-Spitzer satellite, the
ESA-Herschel mission, the NASA-James Webb Space Telescope (JWST), and
many others. In this paper, emphasis will be put on the reference SEDs
of calibration sources for instruments covering the 2 to 200\,$\mu$m
wavelength range. When selecting calibration sources, issues like
brightness in the wavelength regime covered by the instrument,
confusion by neighbouring sources, sky visibility, and pointing
accuracy are to be considered. In the 2--200\,$\mu$m wavelength range,
planets, asteroids, and cool standard stars are the ideal calibration
sources to cover the whole dynamic range of present-day instruments
(see Sect.~\ref{fluxcalibrationsources}). This paper focuses on the
reference SEDs of standard {\em stellar} candles, with the aim of
assessing the reliability and achieving accuracy when using IR stellar
standards for spectral calibration purposes. Quite often, calibration
scientists or observers (have to) rely on `quoted' errors, without
having insight into this part of the calibration pedigree. Since for a
proper interpretation of the output signal one needs to have a grip on
the uncertainties that propagate to the fluxes ultimately calculated
by an instrument, the discussion in this paper is required.


Section~\ref{fluxcalibrationsources} gives an overview of the different
IR flux calibration sources. The status of {\em stellar} reference
SEDs in the IR is discussed in Sect.~\ref{stellarSEDs}, and we focus
on the theoretical atmosphere spectra in Sect.~\ref{status}. Different
stellar reference SEDs are compared in Sect.~\ref{comparison}, and we
end with the conclusions in Sect.~\ref{conclusions}.


\section{Infrared flux calibration sources} \label{fluxcalibrationsources}

In this section, we discuss different IR calibration sources used in
the spectrophotometric calibration process.

\subsection{Planets} \label{planets}

For calibration purposes at wavelengths longer than $\sim$30\,$\mu$m,
planets are extremely useful since they are among the few astronomical
objects bright enough in the far-IR to allow sufficiently accurate
flux density predictions. At 100\,\mic, the flux values for Neptune
and Uranus range between 100 and 1000\,Jy. Uranus is often used as
the primary calibrator, as it is known to be a reliable calibrator
\citep{Griffin1993Icar..105..537G, Sidher2003clim.conf..153S}, while Neptune
and Mars are used as secondary calibrators.

Good models for Uranus exist \citep[e.g.,][]{Moreno1998,
  Orton2003clim.conf..147O, Sidher2003clim.conf..153S}. Radiative
  transfer calculations are done in spherical geometry and take the
  limb into account. The models do not, however, take chemical reactions
  into account, nor heating and cooling. In the mm and submm-regimes,
  the models are confirmed within the accuracy of 10 -- 20\,\%. The
  big advantage of Mars over Uranus is that it is bright. Different
  models for Mars' surface exist
  \citep[e.g.][]{Rudy1987Icar...71..159R, Lellouch2000P&SS...48.1393L,
  Hartogh2005JGRE..11011008H}. Using Mars as a primary calibrator,
  however, would pose different problems: (1) Mars is a planet with a
  surface and an atmosphere, both contributing to the continuum
  emission, (2) sandstorms influence the line shapes, (3) the ice caps
  can influence the continuum if the ice is melting. Neptune could
  serve almost equally well as an independent (secondary)
  calibrator. However, many large features are visible in the Voyager
  IRIS spectra, which may indicate systematic variability with
  rotational phase \citep{Bishop1998P&SS...46....1B}.

\subsection{Asteroids} \label{asteroids}

Typical surface temperatures of main-belt asteroids are such that they
emit the bulk of their thermal radiation in the far-IR. At wavelengths
longer than $\sim$20\,$\mu$m, the largest asteroids are brighter than
the brightest IR stellar sources: they cover the flux range between
100 and 1000\,Jy in the wavelength range between 30 and 45\,\mic, and
hence fill a gap where stellar calibrators are not available
\citep[see Fig.~1 in ][]{Muller1998A&A...338..340M}.

The work of M\"uller \citep{Muller1998A&A...338..340M,
  Muller2002A&A...381..324M, Muller2003clim.conf..157M,
  Muller2005dmu..conf..471M} has largely improved the accuracy of the
  theoretical model SEDs of asteroids. The most accurate SEDs
  at present are based on the thermophysical model (TPM) for describing
  the asteroids' thermal emission
  \citep{Muller1998A&A...338..340M}. Due to the asteroids' orbit,
  brightness temperatures vary by $17 \pm
  13$\,\%. \citet{Muller2002A&A...381..324M} quote an accuracy of
  5\,\% between 5 and 200\,$\mu$m for Ceres, Pallas, and Vesta, i.e.\
  for the three largest asteroids.

The main limitations for asteroids as IR calibration sources come from
 the changing background conditions and the flux changes on
timescales of hours due to rotation. Moreover, the list of reliable
asteroids is still short.

\subsection{Stars} \label{stars}

In the near-IR, stars are ideal calibrators, especially since they are
almost point-like and span a range in flux level more than 4 orders
of magnitude between 2 and 50\,\mic, and one can create a database
covering the whole sky. Stellar standards are, however, quite faint in
the far-IR and can henceforth not be used as primary calibrators at
the far-IR wavelength ranges. Good IR calibration sources need to
comply several criteria: be single and non-variable, not have
an IR excess due to a chromosphere, debris disk, or a circumstellar
envelope, and be located in an uncrowded region that can be
observed all or most of the time \citep{DecinBauwens2006}. Preferably,
the IR standard is cooler than $\sim 10\,000$\,K to provide a good
signal-to-noise ratio over most of the wavelength range.

In the case of \emph{spectrophotometric} calibration, one extra
requirement should be added to this list regarding the composition of
the sample of used calibrators with regard to the spectral types.  For
\emph{photometric} calibration purposes, the use of only one spectral
type can be justified \citep{DiazMiller2007}, but in the case of a
spectrometric calibration, where every single wavelength needs to be
calibrated, none of the available reference SEDs of any spectral type
has high enough accuracy in the full 2 to 200\,$\mu$m wavelength
range (see Sect~\ref{stellarSEDs}).  Three classes of spectral stellar
standards have been commonly used in the IR spectrometric calibration
pedigree: \emph{(i)} early A dwarfs, \emph{(ii)} solar analogs, and
\emph{(iii)} late-type giants, usually of spectral type G9 --
M0\,III. Each of these groups provides a different challenge (see
below), but by combining them, they \emph{(i)} reduce the chances for
systematic errors, possibly introduced by the use of only one spectral
type, and \emph{(ii)} they will increase the total accuracy since the
best SED can be used for each part of the spectrum.


\section{Status of stellar reference SEDs in the IR} \label{stellarSEDs}

\subsection{Black body}

Planck curves represent the simplest way to model stellar far-IR
fluxes. A comparison with the predictions of a solar continuum model
shows clear deviations, arising to more than 20\,\% in the far-IR
\citep[see Fig.~14 in][]{vanderBliek1996A&A...309..849V}. The reason
is the contribution of the H$^-$ free-free opacity, shifting the
flux-forming region more to outer cooler layers. Neither line or
continuous absorption nor back-warming effects are/can be included.

\subsection{Engelke function} \label{Engelke}

The Engelke function \citep{Engelke1992AJ....104.1248E} is a more
sophisticated two-parameter analytical approximation to the 2 --
60\,\mic\ infrared continuum spectrum for giants and dwarfs with
effective temperature, \Teff, between 3500 and 6000\,K. This spectral
function is based on the scaling of a semi-empirical (plane-parallel)
solar atmospheric profile to differing effective temperatures. The
result describes the \emph{continuum} spectrum expected for stars in
terms of their effective temperature and angular size.  The estimated
probable error in absolute flux values is quoted to be $\sim$3\,\%
below 10\,\mic, growing to $\sim$5\,\% around 25\,\mic\ and 6\,\% at
60\,\mic.

\begin{figure}[!thp]
\resizebox{.45\textwidth}{!}{\rotatebox{90}{\includegraphics{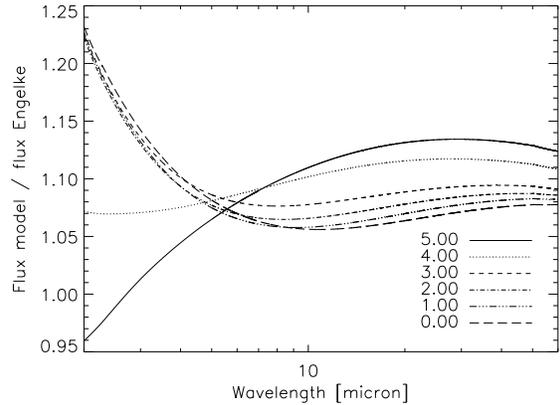}}}
  \caption{Ratio between the continuum flux predicted
  from a theoretical plane-parallel stellar atmosphere with
  \Teff\,=\,3500\,K and values for the logarithm of the gravity ranging
  between 5.00 and 0.00 and the flux computed from the Engelke
  function.
\label{Engelke_PP}}
\end{figure}

The main limitation of the Engelke function is neglect of the
influence of the surface gravity. In Fig.~\ref{Engelke_PP} the
difference between the Engelke function and the continuum flux of a
plane-parallel theoretical stellar atmosphere (see Sect.~\ref{status})
with an effective temperature, \Teff, of 3500\,K for different values
of the gravity is shown. For values of the logarithm of the gravity,
$\log$ g, lower than 3.00, the absolute deviations may be as high as
25\,\%. At a temperature of 6000\,K, the absolute deviations
approximate the errors quoted by
\citet{Engelke1992AJ....104.1248E}. Important, however, for the
discussion in this paper of the RSRF determination are the
\emph{systematic} differences between the Engelke continuum and both
(1) the continua calculated by ab-initio model atmosphere calculations
(see Fig.~\ref{Engelke_PP}) and (2) real continua deduced from
observations \citep[see][]{Engelke2006AJ....132.1445E}. This systematic
discrepancy occurs both for solar analogs and the cooler K--M giants:
the same type of curves as displayed for the lower gravity values in
Fig.~\ref{Engelke_PP} (M-giants) also occur in an analogous plot
for G dwarfs, but with a smaller amplitude of $\sim 3\,\%$. Note
also that for almost all values of $\log$ g, the ratio as displayed in
Fig.~\ref{Engelke_PP} is systematically higher than one, implying that
when either the effective temperature or the angular diameter (\ad) is
kept fixed, the other parameter will be overestimated.

\begin{figure}[!thp]
\resizebox{.45\textwidth}{!}{\rotatebox{90}{\includegraphics{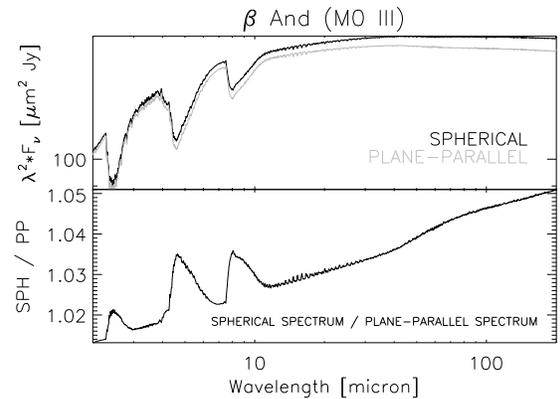}}}
  \caption{{\em Upper panel:} Model atmosphere spectrum in spherical
  geometry (black) compared to the theoretical spectrum calculated in
  plane-parallel geometry (grey) representing the M0~III giant
  \object{$\beta$ And} ($\lambda / \Delta \lambda = 100$). {\em Bottom
  panel:} Ratio between spherical and plane-parallel spectrum. Stellar
  parameters are taken from \citet{Decin2000d}.
\label{PPversusSPH}}
\end{figure}

Another issue not taken into account are sphericity effects, important
when dealing with the extended atmospheres of giants.
Figure~\ref{PPversusSPH} compares a theoretical atmosphere spectrum (see
Sect.~\ref{status}) calculated using spherical geometry with the
spectrum obtained in plane-parallel geometry for the case of an M0~III
giant. Compared to a plane-parallel atmosphere, the radiation field of
a spherical model is diluted in the upper photospheric layers, causing
the temperature (and hence the source function) there to be lower.
The lower surface flux is, however, compensated by the fact that the
IR flux arises from higher layers. The net result is an infrared
excess of the spherical model relative to the plane-parallel
model. Important for this paper is the rising continuum in the bottom
panel of Fig.~\ref{PPversusSPH}, implying that the use of the Engelke
function or a plane-parallel geometry to represent the SED of giants
may introduce uncertainties on the broad-band RSRF characterisation of
a few percent.

\subsection{The IR calibration framework by Cohen et al.} \label{Cohen}

In a series of papers on spectral irradiance calibration in the
infrared, Cohen and colleagues have presented a self-consistent
network of absolutely calibrated reference spectra between 1.2 and
35\,\mic\ for over 600 stars spanning spectral types from A0 to M4 and
distributed over the entire sky \citep[][often referred to as the CWW
network]{Cohen1992AJ....104.1650C, Cohen1992AJ....104.2030C,
Cohen1995AJ....110..275C, Cohen1996AJ....112..241C,
Cohen1999AJ....117.1864C, Cohen2003AJ....125.2645C}. The CWW network
is the result of a tremendous and consistent work done during the past
15 years.  Four main steps can be distinguished in the network. (i)
The approach is based on a pair of absolutely calibrated models of the
two A-type dwarfs \object{Vega} and \object{Sirius} calculated by
\citet{Kurucz1993KurCD..13.....K}
\citep{Cohen1992AJ....104.1650C}. (ii) The next layer in the network
is a set of \emph{composites} of 13 secondary calibration
stars. Composites are spectra measured by various ground-based and
airborne telescopes (KAO, IRAS LRS, etc.) that are averaged and/or
spliced together to form a continuous spectrum between 1.2 and
35\,\mic. This set compromises one G2 dwarf --- \object{$\alpha^1$
Cen} (G2~V); seven K giants --- \object{$\alpha$ Tau} (K5~III),
\object{$\alpha$ Boo} (K2~IIIp), \object{$\beta$ Gem}(K0~III),
\object{$\alpha$ Hya} (K3~II-III), \object{$\alpha$ TrA} (K2~III),
\object{$\epsilon$ Car} (K3~III), \object{$\gamma$ Dra} (K5~III); and
5 M giants --- \object{$\beta$ Peg} (M2.5~II-III), \object{$\beta$
And} (M0~III), \object{$\alpha$ Cet} (M1.5~III), \object{$\gamma$ Cru}
(M3.4~III), \object{$\mu$ UMa} (M0~III)
\citep{Cohen1992AJ....104.2030C, Cohen1995AJ....110..275C,
Cohen1996AJ....112..241C, Cohen1996AJ....112.2274C}. The KAO was no
longer operational at the time the composites of \object{$\gamma$
Dra}, \object{$\alpha$ Cet}, \object{$\gamma$ Cru} and \object{$\mu$
UMa} were constructed.  Spectral regions that are opaque from the
ground were therefore replaced by spectra of other stars with the same
spectral type. Note that the `composite' of \object{$\alpha^1$ Cen} is
a Kurucz model-atmosphere spectrum, which was then used as a reference
to create the composite spectra of \object{$\alpha$ TrA} and
\object{$\epsilon$ Car}. For wavelengths longer than $\sim$22\,\mic,
the Engelke function was used for twelve composites to represent the
stellar spectrum, the exception being $\alpha$ Tau. (iii) These 13
composite spectra are the basis for a set of absolutely calibrated
spectral \emph{templates} for 602 stars with spectral types between
G9.5-M0.5~III \citep{Cohen1999AJ....117.1864C,
Cohen2003AJ....125.2645C}. For each star a smoothed composite spectrum
is chosen as its spectral template according to the spectral
class. The template is corrected for reddening and normalised using
available optical photometry. For purpose of the NASA-Spitzer Infrared
Array Camera (IRAC) calibration an extra set of absolutely calibrated
$0.275 - 35$\,\mic\ spectra of 33 optical standard stars with spectral
types between A0 -- A5~V and K0 -- M0~III, and flux levels down to $V
\sim 11-12$, was prepared \citep{Cohen2003AJ....125.2645C}. (iv) In
support of the Infrared Space Observatory (ISO) PHT and LWS
instruments, CWW have provided extrapolated continuum spectra of few
composites till 300\,\mic\ \citep{Cohen1996AJ....112.2274C}. The
continuum spectra have been interpolated in a grid of four
\emph{plane-parallel} model-atmosphere spectra (see also
Sect.~\ref{Engelke}) with solar abundances. To obtain the
temperature-versus-continuum optical depth relation
($\tau_{\rm{cont}}$), a scaling based on the ratio
\Teff(desired)/\Teff(grid model) was performed. Differences in gravity
were not accounted for. An uncertainty of 5.67\,\% was assessed to
these extrapolations.

It should be emphasised that the CWW network has some  major
  advantages, the main ones being \emph{(1)} the common calibration
  pedigree, and \emph{(2)} the good sky-coverage \citep[see Fig.~2
  in][]{Cohen1999AJ....117.1864C}.   For that reason, the CWW
  network was used to support the calibration of many spaceborne,
  airborne, and ground-based instruments, as e.g. instruments onboard
  ISO, Spitzer, AKARI, MSX, etc.

 However, aiming for a (spectrophotometric) calibration better
than a few percent, some steps used in the construction of the CWW
network deserve some comments.

\begin{itemize}
\item{A first issue concerns the danger of propagating of systematic
  errors from the top of the calibration pedigree. In this context, we
  want to note that the adopted spectra for the fundamental reference
  stars in the calibration pedigree are absolutely-calibrated
  \emph{Kurucz models} of the A-type stars \object{Sirius} and
  \object{Vega}. Since no absolute flux measurements were available
  for \object{Sirius}, \citet{Cohen1992AJ....104.1650C} chose to
  calibrate the \object{Sirius} model absolutely with respect to
  \object{Vega} by using the observed magnitude differences in $K$,
  $L$, $L^{\prime}$, $M$, and at four mid-IR wavelengths.  However,
  the use of Vega (or its model spectrum representing an `ideal' Vega
  without flux excess) is strongly debated \citep[for an overview, we
  refer to][]{Gray2007ASPC..364..305G}. We summarise that (i) it is
  obvious that the \object{Vega} debris disk system contributes to the
  flux excess from this star beyond 10\,\mic\
  \citep[e.g.][]{Neugebauer1984Sci...224...14N}, (ii) that recent
  results by \citet{Absil2006A&A...452..237A} show an excess of $1.29
  \pm 0.19$\,\% in the K-band close-in to Vega (within 1'') presumably
  due to a hot inner circumstellar disk; (iii) and \object{Vega} is a
  rapidly rotating pole-on star with a significant temperature
  gradient ($\sim$1500 -- 2250\,K) from its pole to its equator
  \citep{Gulliver1994ApJ...429L..81G, Aufdenberg2006astro.ph..3327A},
  so its SED differs from conventional models, not assuming a rotating
  star \citep{Gray2007ASPC..364..305G}.  As noted by
  \citet{Bohlin2007ASPC..364..315B} a single-temperature
  \Teff\,=\,9400\,K Kurucz model fits the Hubble Space Telescope (HST)
  STIS SED to 1-2\,\% from 3200 to 10000\,\AA. However, a
  multi-temperature model may be required to achieve a $\sim$1\,\%
  precision for shorter and longer wavelengths (the topic of this
  study).  Finally (iv) previous work has suggested an inconsistency between
  infrared and visible measurements of \object{Vega}, when fitted to a
  standard A0~V star model \citep[e.g.][]{Rieke1985AJ.....90..900R,
  Megessier1995A&A...296..771M, Verhoelst2005PhDT.........2V,
  Rieke2007}, probably having its roots in \object{Vega} being a
  rapidly rotating star seen pole-on. When the absolute calibration of
  the \object{Sirius} spectrum by magnitude-difference with
  \object{Vega} is done in the optical and not in the IR, the
  fluxes of the \object{Sirius} spectrum increases by
  $\sim$3.7\,\% \citep{Verhoelst2005PhDT.........2V, Rieke2007}
  compared to the absolute calibration as proposed by
  \citet{Cohen1992AJ....104.1650C}, shifting the angular diameter of
  Sirius from 6.04\,mas \citep{Cohen1992AJ....104.1650C} to 6.15\,mas
  (see Fig.~\ref{FigTijl}). Based on the \emph{Midcourse Space
  Experiment} (MSX) data, \citet{Price2004AJ....128..889P} propose an
  increase in the SED of \object{Sirius} by 1\,\%. This increase in
  the flux density of the primary standard \object{Sirius} has a
  direct influence on the absolute flux level of all composites and
  templates, whose absolute photometric data are based on relative
  flux differences compared to the primary standards.

\begin{figure}[!thp]
\resizebox{.45\textwidth}{!}{\rotatebox{0}{\includegraphics{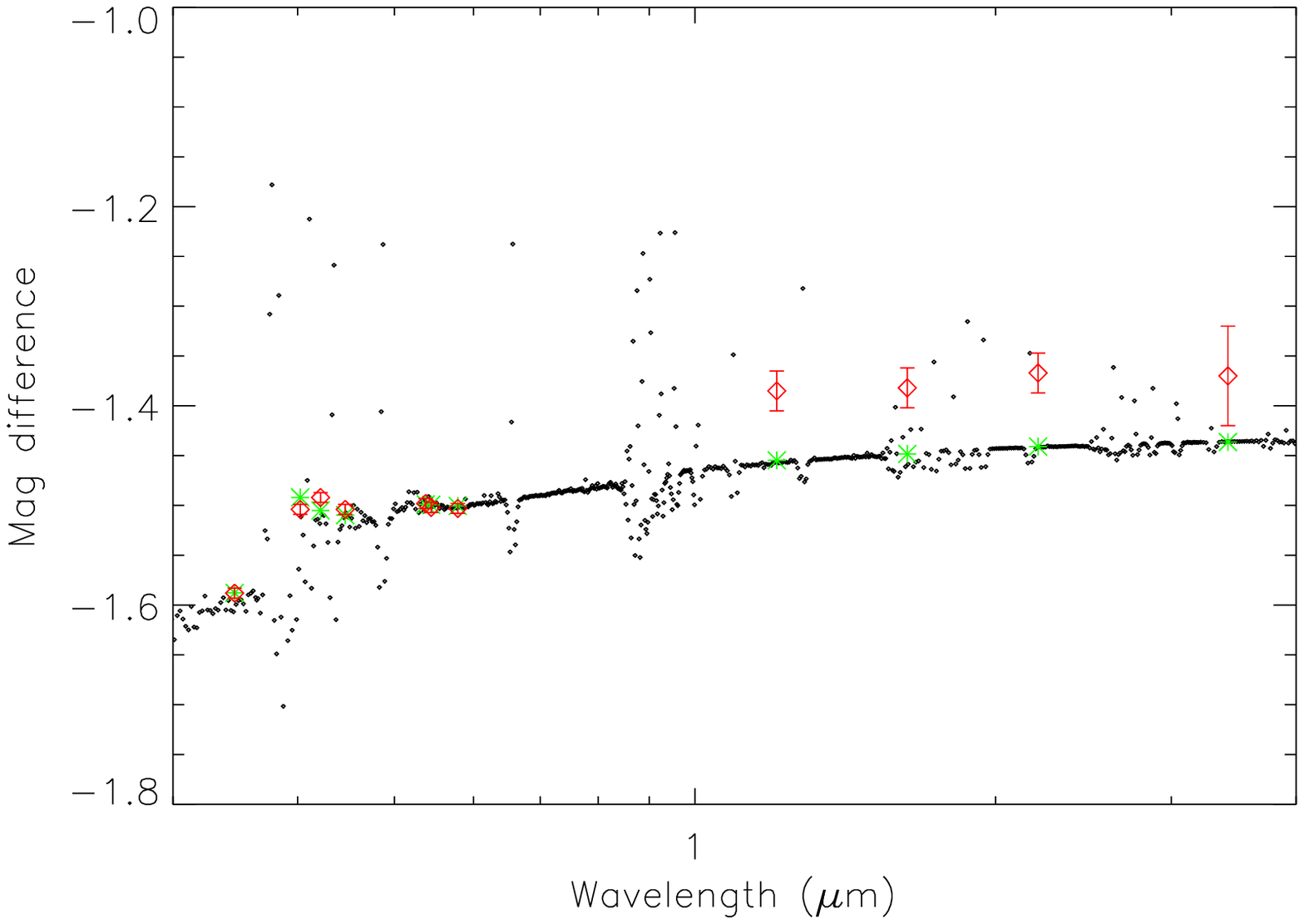}}}
  \caption{ Observed (red - diamonds) versus synthetic (green -
  stars) magnitude difference between Vega and Sirius
  (Mag$_{\rm{Vega}}$ $-$ Mag$_{\rm{Sirius}}$) in the Geneva
  photometric system and the JHKL bands
  \citep{Verhoelst2005PhDT.........2V}. Black dotted line represents
  the difference between both of Kurucz's models used in
  \citet{Cohen1992AJ....104.1650C}, with stellar parameters 
  \Teff\,=\,9550\,K, $\log$\,g\,=\,3.95, [Fe/H]\,=\,$-0.3$ for Vega
  and \Teff\,=\,9850\,K, $\log$\,g\,=\,4.25, [Fe/H]\,=\,$+0.5$ for
  Sirius. However, the Kurucz's model for Sirius is scaled in such a
  way as to match the observed difference in the Geneva photometric
  system, yielding an angular diameter for Sirius of 6.15\,mas
  \citep[instead of 6.04\,mas as used
  in][]{Cohen1992AJ....104.1650C}. Clearly, the (N)IR magnitude
  differences are not compatible asthey differ by up 10\,\%.
\label{FigTijl}}
\end{figure}

}
\item{ One should be aware that, for all composite
  spectra, with the exception of $\alpha$ Tau, the flux values between
  $\sim 22$ and 35\,\mic\ are represented by the Engelke function. As
  discussed in the previous section, the neglect of both gravity and
  sphericity effects results in a systematic broad-band inaccuracy of
  $\sim 3\,\%$ for early A dwarfs, rising to $\sim 8\,\%$ for late
  K-giants. \citet{Cohen1996AJ....112.2274C} and
  \citet{Cohen2007ASPC..364..333C} justify this assumption by
  comparing the UKIRT CGS3 16 -- 24\,\mic\ spectra of \object{$\alpha$
  Boo} and \object{$\alpha$ Cet} with their relevant Engelke
  approximation \citep[see Fig.~5 in][]{Cohen1996AJ....112.2274C} and
  from the KAO spectrum of \object{$\alpha$ Tau}
  \citep{Cohen1992AJ....104.2030C}. A comparison to these observations
  in the $\lambda^4 F_{\lambda}$-space (Fig.~\ref{fig2}) does,
  however, not give any more support to the ``flatness'' of an Engelke
  spectrum than to the rising continuum of a theoretical model
  atmosphere (see also Sect.~\ref{comparison}).  One can only argue
  that, from a physical point of view, the increase in the H$^{-}$
  opacity towards longer wavelengths ($\sigma({\rm{H^{-}_{ff}}})
  \propto \lambda^2$) yields a shift of the flux-forming region
  towards higher, cooler, atmospheric layers, and hence a decrease in
  the stellar surface flux.

\begin{figure}[!thp]
\begin{center}
\resizebox{.43\textwidth}{!}{\rotatebox{90}{\includegraphics{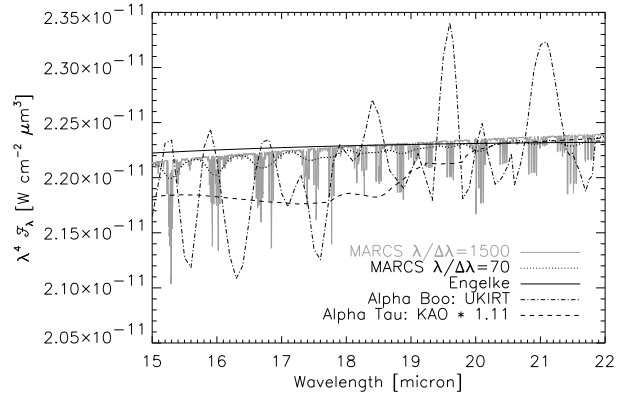}}}
  \caption{UKIRT CGS3 15 -- 22\,\mic\ spectrum for \object{$\alpha$
  Boo} (K2~IIIp; unscaled, dash-dot line) and KAO spectrum for
  \object{$\alpha$ Tau} (K5~III; scaled by a factor 1.11, dashed line)
  are compared with the Engelke approximation (full line) with
  parameters as used by \citet{Cohen1996AJ....112.2274C} for
  \object{$\alpha$ Boo} (\Teff\,=\,4362\,K, $\theta_d = 21.12$\,mas),
  and an appropriate theoretical {\sc marcs} model at a resolution
  $\lambda / \Delta \lambda = 1500$ (grey line) and at a resolution of
  $\lambda / \Delta \lambda = 70$ (dotted line; corresponding to the UKIRT CGS3
  resolution), with stellar parameters as determined in
  \citet{Decin2000d}.
\label{fig2}}
\end{center}
\end{figure}

}

\item{The assumption in the creation of a `template' spectrum is that
  a de-reddened spectrum of a star is representative of the infrared
  spectral signature of any star of the same two-dimensional spectral
  classification. This is only true to some extent. Differences in
  stellar parameters and photospheric abundances can result in
  substantial differences in the strength of molecular absorption
  bands as can be seen in Fig.~\ref{compparam}. A more elaborated
  version of this figure can be found in
  \citet{Decin2000A&A...364..137D}. For G-type stars and cooler,
  differences in stellar parameters within one spectral 
  sub-class (e.g.\ from K1\,III to K2\,III) may cause
  differences in the (broad) NIR molecular absorption bands in the
  order of 5 -- 8\,\% (at a resolution of 100), an uncertainty that
  should be directly translated to the uncertainties attributed to the
  \emph{template} spectra.  For hotter stars, the deviations are
  smaller and are mainly seen in the strength of the hydrogen
  lines. The continuum flux may differ by up to $\sim$4\,\%.}
\item{Recently, \citet{Engelke2006AJ....132.1445E} found, for
  a subset of 33 standards with spectral types between G2 and M5,
  available in the CWW network, that the derived spectra  (by Engelke
  and co-workers) display systematic differences with the
  composites/templates in the CWW network, in that Engelke
  standard spectra have 4--7\,\% lower flux in the 1--4\,\mic\
  spectral region. The spectra of \citet{Engelke2006AJ....132.1445E}
  remove the A--K star calibration bias recently noted in the
  calibration of Spitzer/IRAC, which was based on templates in the CWW
  network \citep{Reach2005PASP..117..978R}.}
\end{itemize}

\begin{center}
\begin{table*}[htp!]
\caption{\label{overview_uncertainties_cohen}
 Summary of uncertainties attributed to the standard star spectra in
the CWW network in the 2--200\,\mic\ range.}
\begin{tabular}{llll}\hline \hline
  \rule[-3mm]{0mm}{8mm}Description & Uncertainty & Spectral Type
  & Wavelength Region \\ 

\hline \rule[-0mm]{0mm}{5mm}$\bullet$ propagation of systematic
  errors & & & \\

\phantom{$\bullet$ } e.g.\ due to use of Vega as prime calibrator & &
  & \\

\phantom{$\bullet$ } $\rightarrow$ absolute flux calibration of Sirius
   w.r.t.\ Vega & 1--3\,\% & A--M & 2--200\,\mic \\

$\bullet$ use of Engelke function \citep{Engelke1992AJ....104.1248E} &
3\,\% & A-dwarfs & 22--35\,\mic \\

 & 8\,\% & K-giants & 22--35\,\mic \\

$\bullet$ creation of `template' spectrum & & & \\

\phantom{$\bullet$ }$\rightarrow$ molecular features & up to 8\,\% &
  G--K & around 2.3, 4.0, 4.2, 8\,\mic \\

\phantom{$\bullet$ }$\rightarrow$ continuum & up to 4\,\% & A--M &
  2--200\,\mic \\

$\bullet$ presence of chromosphere/ionised wind & $\ga$10\,\% & G--M &
$\lambda > 100$\,\mic \\

\phantom{$\bullet$ }presence of circumstellar dust & $\ga$10\,\% &
A--M & $\lambda > 2$\,\mic \\

$\bullet$ systematic difference with new Engelke & & & \\

\rule[-3mm]{0mm}{3mm}\phantom{$\bullet$ } standards \citep{Engelke2006AJ....132.1445E} &
  \raisebox{1.5ex}[0pt]{4--7\,\% } & \raisebox{1.5ex}[0pt]{G2--M5 } &
  \raisebox{1.5ex}[0pt]{1--4\,\mic}\\

\hline
  \end{tabular}
\end{table*}
\end{center}

\begin{figure}[!thp]
\begin{center}
\resizebox{.48\textwidth}{!}{\rotatebox{90}{\includegraphics{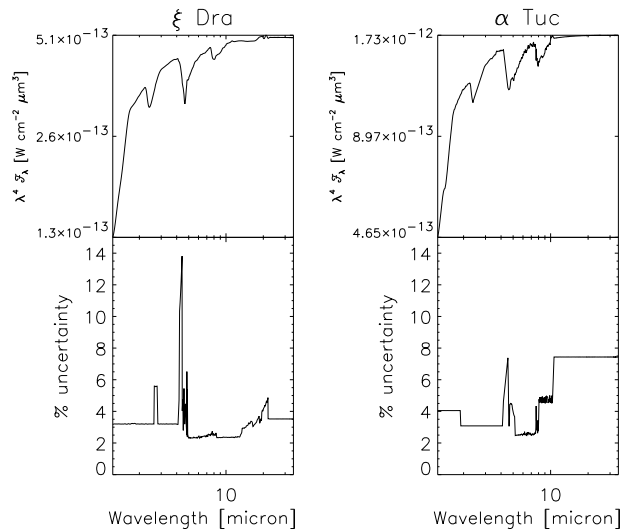}}}
\vspace*{4ex}
  \caption{{\em Upper panels:} \emph{Template} spectra of
  \object{$\xi$ Dra} (K2~III) and \object{$\alpha$ Tuc} (K3~III). The
  composite spectrum of \object{$\alpha$ Boo} (K2~IIIp) has been used
  to construct the template of \object{$\xi$ Dra} (K2~III), and for
  \object{$\alpha$ Tuc} (K3~III) the composite of \object{$\alpha$ Hya}
  (K2~II-III) is used. {\em Lower panels:} the total uncertainty
  attributed to the templated spectra by
  \citet{Cohen1999AJ....117.1864C}. Main absorption features are the
  CO fundamental around 4.5\,$\mu$m, CO first overtone around
  2.3\,$\mu$m, SiO fundamental around 8\,$\mu$m and SiO first overtone
  around 4.2\,$\mu$m.
\label{fig_uncertainty}}
\end{center}
\end{figure}

The net result is that the CWW network has provided the astronomical
  community with a consistent network containing some 600
  stars. Calibration experiments for the SPIRIT~III infrared
  instrument onboard the MSX have demonstrated that in the 6 (broad)
  spectral bands (with wavelength between 4.29 and 21.34\,\mic) the
  \emph{absolute} flux calibration for 8 composites in the CWW network
  is $<2.3$\,\% \citep{Price2004AJ....128..889P}. However, previous
  discussion also shows that for some spectral regions of the IR SED,
  the uncertainties attributed to the template spectra or extrapolated
  continuum spectra in the CWW network should be larger than now
  quoted (for an overview, we refer to
  Table~\ref{overview_uncertainties_cohen}). A typical example of the
  attributed total uncertainty for two template spectra is given in
  Fig.~\ref{fig_uncertainty}. In the spectral regions where molecules
  absorb photons substantially, at least an 8\,\% uncertainty (see above)
  should be attributed to the templated spectra. Another example is
  the (unknown) contribution of a (possible) chromosphere, ionised
  wind or dust-driven wind, which may yield uncertainties in the order
  of 10\,\% at 100\,\mic\ \citep[see
  Sect.~\ref{chromosphere};][]{vanderBliek1996A&A...309..849V}, much
  higher than the 5.67\,\% uncertainty attributed to the extrapolated
  composite spectra by \citet{Cohen1996AJ....112.2274C}.

 In order to remove the `molecular feature' bias that can enter
into the template construction, the computation of ab-initio
theoretical model atmosphere spectra may offer a solution (see next
section). Moreover, theoretical atmosphere spectra can be computed
at any spectral resolution (comparable to that of the instrument) and
in that way compete with the template spectra, which have quite low
spectral resolution.

\subsection{Model atmosphere spectra} \label{theoretical_spectra}

Different groups have given a lot of effort into developing trustworthy
\emph{hydrostatic 1-dimensional} (1-D) model atmosphere codes, the
most famous ones in the field of cool stellar atmospheres (\Teff\ $\la
10\,000$\,K) being the {\sc atlas} atmosphere code
\citep{Kurucz1970SAOSR.308.....K, Kurucz1993KurCD..13.....K,
Kurucz1996ASPC..108..160K, Castelli2004astro.ph..5087C}, the {\sc
marcs} code (originally presented in
\citealt{Gustafsson1975A&A....42..407G}, with the current grid being
described in \citealt{Gustafsson2003ASPC..288..331G}), and the {\sc
phoenix} code \citep{Hauschildt1999ApJ...512..377H,
Hauschildt1999ApJ...525..871H}. Also in the case of late-type (variable)
asymptotic giant branch (AGB) stars, the 1-D hydrostatic models are
still useful in providing us with physical information, due to the
great detail now included in their treatment of
opacities. \emph{Hydrodynamic} models in 1-D
\citep[e.g.][]{Bessell1996A&A...307..481B, Winters2000A&A...361..641W,
Hofner2003A&A...399..589H} include detailed physics, like
time-dependent dust formation and the coupling between gas, dust, and
radiation. Few dynamical models are now able to simultaneously solve
the equations of hydrodynamcics and frequency-dependent radiative
transfer leading to consistent dynamical density-temperature
structures \citep[see, e.g.,][]{Hofner2003A&A...399..589H}. Using these
kinds of models, one can already quantitatively reproduce the variation
in line profiles due to the influence of gas velocities \citep[see,
e.g.,][]{Nowotny2005A&A...437..273N}. Recently, \emph{3-D
hydrodynamical} simulations of stellar surface convection have become
feasible thanks to advances in computer technology and efficient
numerical algorithms \citep{Collet2006ApJ...644L.121C}. The 3-D models
can shed light on the coupling between convection and pulsation, as
well as be employed in element abundance analysis. These 3-D model
computations are, however, still very computer-time intensive, and it is
not known how well they represent the surface layers of the stars, so
spectrum synthesis of a whole sample of stellar spectra is still
beyond the scope.

For the scope of this project of developing accurate SEDs for IR stellar
standards, we therefore rely on a 1-D hydrostatic model
atmosphere code. We here opt to use the {\sc marcs} code (see next
section).

\section{Status and accuracy of {\sc marcs} model atmosphere spectra in the IR}
\label{status}

The {\sc marcs} model atmosphere and spectrum synthesis code was
developed in Uppsala
\citep{Gustafsson2003ASPC..288..331G}{\footnote{http://marcs.astro.uu.se}}.
This local thermodynamic equilibrium (LTE) model atmosphere code is
built on the assumptions of spherical or plane-parallel stratification
in homogeneous stationary layers and hydrostatic equilibrium. Energy
conservation is required for radiative and convective flux, where the
energy transport due to convection was treated through a local
mixing-length theory. For a discussion of the method for solving the
radiative transfer equations in the atmospheric models and spectrum
synthesis, we refer to \citet{Nordlund1984mrt..book..211N} and
\citet{Plez1992A&A...256..551P}.

\begin{figure}[!thp]
\begin{center}
\resizebox{.48\textwidth}{!}{\rotatebox{90}{\includegraphics{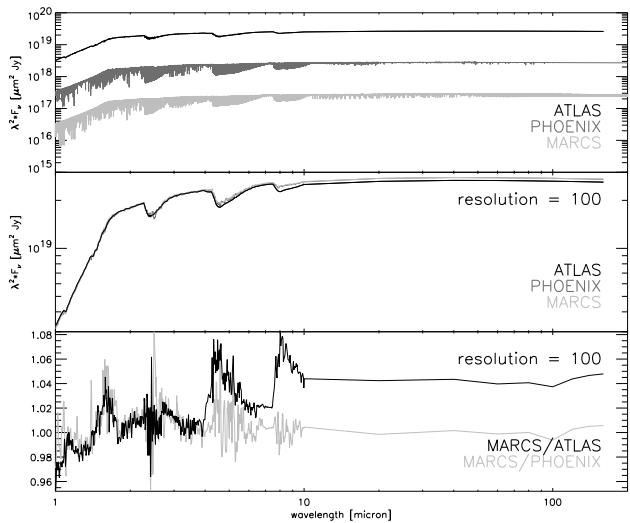}}}
\vspace*{4ex}
  \caption{ {\em Upper panel:} Comparison between the {\sc atlas},
  {\sc phoenix}, and {\sc marcs} model spectra. The {\sc phoenix}
  model is shifted downwards with a factor 10, and the {\sc marcs}
  model with a factor 100. {\em Middle panel:} Comparison between the
  {\sc atlas}, {\sc phoenix} and {\sc marcs} model spectra at a
  resolution of 100 for $\lambda \le 10$\,\mic. At longer wavelengths,
  only the 8 wavelength points as given in the {\sc atlas} models (at
  20, 40, 60, 80, 100, 120, 140, and 160\,\mic) are
  retained. Difference between {\sc phoenix} and {\sc marcs} model
  fluxes is barely visible. {\em Bottom panel:} Ratio between {\sc
  marcs} and {\sc atlas} (black) and {\sc phoenix} (gray) model
  fluxes.
\label{fig_compare_models}}
\end{center}
\end{figure}

\begin{figure*}[!thp]
\begin{minipage}[t][21truecm][t]{\textwidth}
  \sidecaption
  \includegraphics[width=12cm,angle=0]{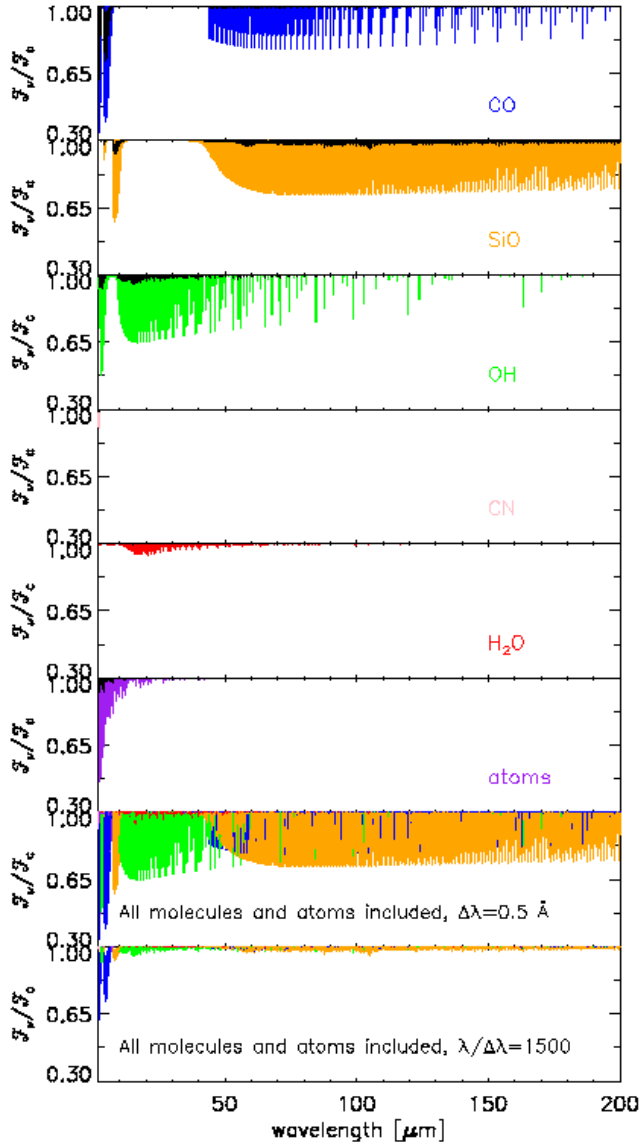}
  \caption{Contribution of CO (blue), SiO (orange), OH (green), CN
  (pink), H$_2$O (red), atoms (purple) to the $2-200$\,$\mu$m
  synthetic spectrum calculated from the atmosphere model with stellar
  parameters \Teff\,=\, 4320\,K, $\log$ g [cm/s$^2$] \,=\, 1.50, mass
  M = 1.1\,\Msun, [Fe/H]\,=\,$-0.50$, microturbulence $\xi_t =
  2$\,\kms, $\varepsilon$(C)\,=\,7.96, $\varepsilon$(N)\,=\,7.61,
  $\varepsilon$(O)\,=\,8.68, $\varepsilon$(Mg)\,=\,7.33,
  $\varepsilon$(Si)\,=\,7.20, and $^{12}$C/$^{13}$C\,=\,7. The first
  six panels display the contribution at a resolution $\Delta \lambda
  = 0.5$\,\AA\ in color and at a medium resolution $\lambda/\Delta
  \lambda = 1500$ in black. The total contribution to the full
  spectrum is shown in the seventh panel at the high resolution of
  $\Delta \lambda = 0.5$\,\AA, and in the bottom panel at the medium
  resolution of $\lambda/\Delta \lambda = 1500$. Note the lack of
  atomic absorption features for wavelengths longer than
  $\sim$50\,$\mu$m.}
  \label{abooFIR} 
\end{minipage}
\end{figure*}

 In the framework of the release of a grid of {\sc marcs} models,
  colleagues from the University of Uppsala are comparing the model
  structures of the {\sc marcs}, {\sc atlas}, and {\sc phoenix}
  codes. An example typical of a cool K giant is shown in
  Fig.~\ref{fig_compare_models}. The plot is made in the $\lambda^2
  F_{\nu}$-space to clearly distinguish the differences. The continuum
  level between the {\sc marcs} and {\sc phoenix} models do agree
  within 0.5\,\%.  A larger difference of $\sim$5\,\% is seen between
  the {\sc marcs} and {\sc atlas} models for the continuum at $\lambda
  > 10$\,\mic. The difference in the molecular features is $\la$4\,\%
  between {\sc marcs} and {\sc phoenix} and $\la$8\,\% between {\sc
  marcs} and {\sc atlas} at a resolution of 100. The reason for the
  difference is the use of a plane-parallel geometry in the case of the
  {\sc atlas} models, whereas the {\sc phoenix} and {\sc marcs} models
  use a spherically symmetric geometry. Slightly different surface
  temperature and geometrical extension explain the difference between
  {\sc marcs} and {\sc phoenix} model fluxes. Moreover, the {\sc
  marcs} models are based on the \citet{Asplund2004A&A...417..751A}
  solar abundances, while the {\sc phoenix} models follow the results
  of \citet{Grevesse1996ASPC...99..117G}.  Note also that the {\sc
  marcs} model uses a more complete OH and SiO line list at $\lambda >
  10$\,\mic\ (see also Fig.~\ref{abooFIR}).

For spectrum synthesis, data on the absorption by atomic and molecular
species are collected from different databases. For a discussion of
various available atomic and molecular line lists, we refer to
\citet{Decinthesis}. For the purpose of this work, the following infrared
spectroscopic line lists were used: CO line list computed by
\citet{Goorvitch1994ApJS...91..483G}, SiO by \citet{Langhoff1993}, CN
by Plez ({\em priv.\ comm.}), OH by \citet{Goldman1998}, H$_2$O by
\citet{PartridgeSchwenke1997}, NO, HF, NH, HCl, and CH by Sauval ({\em
priv.\ comm.}), and atomic line lists by
\citet{Vanhoof1998sese.conf...67V}, by \citet{Hirhor95}, by Sauval
({\em priv.\ comm.}), and of VALD \citep{Piskunov1995lahr.conf..610P,
Ryabchikova1997BaltA...6..244R,Kupka1999A&AS..138..119K}.  Using the
model photosphere as input, synthetic spectra are calculated for a
typical resolution of $\lambda/\Delta \lambda \sim 300\,000$, even
though the final instrumental resolution is often lower. With a
typical microturbulence of 2\,\kms, this means we are certain to sample
all lines in the atomic and molecular database. We note that, for the
purpose of calculating theoretical spectra in the mid to far-IR, some
line lists are still far from complete or accurate (see
Sect.~\ref{linelists}). An typical example of a K2 giant is shown in
Fig.~\ref{abooFIR}. At a medium resolution of $\lambda/\Delta \lambda
= 1500$, the depression of flux due to the line veiling is $\sim$3\,\% at
wavelengths longer than $\sim$30\,$\mu$m.

  For the purpose of using model atmosphere spectra to represent the
  SED of standard stars, an appropriate analysis of different sources
  of uncertainties contributing to total uncertainty in the spectrum
  predictions is in place. The following sections discuss the effects
  of sources of error as {\em a)} the input stellar parameters, {\em
  b)} uncertainties in the model assumptions, {\em c)} the possible
  presence of a chromosphere, ionised wind or circumstellar dust
  shell, {\em d)} the adopted continuum opacity, {and \em e)} the used
  line lists. 

\subsection{Dependency on stellar parameters} \label{dependency}

\begin{figure}[!htp]
\includegraphics[width=.35\textwidth,angle=90]{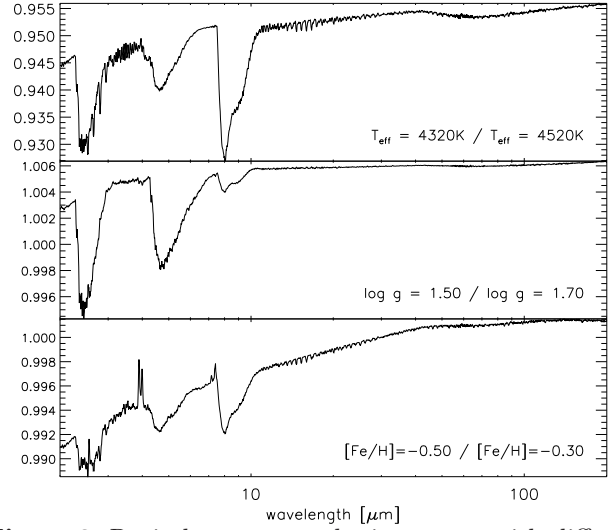}
\vspace*{2.5ex}
\caption{Ratio between synthetic spectra with differing stellar
  parameters at a resolution of $\lambda / \Delta \lambda = 100$. The
  standard model has the same stellar parameters as in
  Fig.~\ref{abooFIR}.  {\em Upper panel:} Ratio between a synthetic
  spectrum with \Teff\,=\,4320\,K and with \Teff\,=\,4520\,K. {\em
  Middle panel:} Ratio between $\log$ g\,=\,1.50 and $\log$
  g\,=\,1.70. {\em Bottom panel:} Ratio between [Fe/H]\,=\,$-0.50$ and
  [Fe/H]\,=\,$-0.30$.}
\label{compparam}
\end{figure}

Model atmospheres are defined by the fundamental parameters effective
temperature, gravity, and metallicity (and stellar mass or radius in
case of a spherical geometry). As demonstrated in
\citet{Decin2000A&A...364..137D}, the influence of the stellar mass on
the synthetic spectrum is small. Varying the other fundamental
parameters shows the effect of errors in the determination of the
fundamental parameters on the synthetic flux distribution. Although
this effect is dependent on the full set of fundamental stellar
parameters, we may summarise that, for $\lambda > 50$\,$\mu$m {\em
(1.)}, varying the effective temperature by $\sim$200\,K for stars
with spectral type between G and K, roughly corresponds to a change of
$\sim$4\,\% in the continuum flux in the IR, {\em (2.)} a change in
the logarithm of the surface gravity of 0.20\,dex introduces
uncertainties in the continuous flux distribution of about 0.5\,\%,
and {\em (3.)}  an uncertainty in the metallicity of about 0.20\,dex
corresponds to an uncertainty in the IR continuum flux of
$\sim$0.1\,\%. The uncertainty in the near-IR and on molecular
absorption features is, however, much greater, as can easily be seen in
Fig.~\ref{compparam}, arising to 7\,\% (at a resolution of 100).

The same order of magnitude on the uncertainties also applies to the
continuum predictions of the hotter A-type stars for the same relative
changes in stellar parameters. On one hand, one may argue that the
absence of broad molecular absorption bands makes these hotter dwarfs
more eligible as standards. On the other hand, exactly the presence of
molecular features in the cooler giants can be used as a strong
diagnostic tool to estimate the stellar parameters, and hence to
provide a fiducial prediction of the whole IR spectrum. 
 Especially in spectral regions with strong hydrogen lines, for
which the computation of the self-broadening remains problematic
\citep{Barklem2000A&A...355L...5B, Decin2000b}, the use of cooler
giant as stellar standards will improve the accuracy of the
spectrophotometric calibration.

\subsection{The temperature distribution $T(\tau)$}

Model atmosphere calculations are based on a number of assumptions,
one of them being radiative and convective flux conservation also in
the outermost layers of the photosphere. This simplification may be
the reason that the, otherwise almost perfect, match between the
FTS-Kitt Peak Spectrum of \object{$\alpha$ Boo} and theoretical
predictions based on {\sc marcs} atmospheres \citep[see Figs.~16-17
in][]{Decin2000b} shows small deviations at the 1--2\,\% level at a
resolution of $\lambda/\Delta \lambda \sim 60\,000$ in the
low-excitation CO and OH lines. 

Not only for K giants (as \object{$\alpha$ Boo}), but also for the
most studied star, the Sun, there is still quite some debate
concerning the assumed mean thermal profile in the outer layers of the
photosphere. Both for G and K-type stars, indications are found from,
e.g., the Ca~II H and K lines that the temperature structure has a
minimum before segueing into the mechanically heated chromosphere
\citep[e.g.][]{Ayres1975ApJ...200..660A}. Controversially, the
analysis of CO lines indicates a cooler brightness temperature at the
same altitude \citep[e.g.][]{Ayres1981ApJ...245.1124A,
Wiedemann1994ApJ...423..806W, Ayres2006ApJS..165..618A}. 

As done by \citet{vanderBliek1996A&A...309..849V} and
\citet{Vanhollebekethesis}, one can simulate uncertainties in the
temperature structure $T(\tau)$ to mimic different effects: {\em a)}
flatten or steepen $T(\tau)$ for a Rosseland optical depth
$\tau_{\rm{ross}} < 0.01$ to study the effect of less, respectively
more, line blanketing, {\em b)} steepen $T(\tau)$ around
$\tau_{\rm{ross}} = 0.1$ to study the effect of convective overshoot
in the outer layers of the model photosphere, or {\em c)} steepen
$T(\tau)$ for $\tau_{\rm{ross}} > 1$ to study the effect of convection
in the deeper layers of the atmosphere. From Table~3 in
\citet{vanderBliek1996A&A...309..849V} and the results of
\citet{Vanhollebekethesis}, we can conclude that this type of error
in the temperature distribution gives rise to uncertainties in the
predicted continuum flux at 100\,$\mu$m lower than 3.5\,\%. This
uncertainty can, however, be strongly reduced in case high-resolution
spectra are available for atomic or molecular lines with different
strengths and excitation energies. From a proper analysis of the line
shapes, one can pin down $T(\tau$) to $\sim$50\,K in the outer layers,
reducing the uncertainties in IR {\em continuum} flux predictions to
$\sim$1--2\,\% \citep{vanderBliek1996A&A...309..849V}.

\subsection{The presence of a chromosphere, ionised wind, or circumstellar dust
  shell} \label{chromosphere}

One of the largest uncertainties in model spectrum predictions is the
(unknown) presence of a chromosphere, an ionised wind, or a
circumstellar dust envelope. These structures may easily yield a flux
excess in the order of 10\,\% at mid to far-infrared wavelengths
\citep{vanderBliek1996A&A...309..849V}.  Recent studies by
\citet{Dehaes2006b} show that all of the six studied cool standard
stars, used for the calibration of the Short-Wavelength Spectrometer
(SWS, 2.38 -- 45.2\,\mic) onboard ISO, with spectral types between
K2~III and M0~III show a flux excess due to an ionized wind. Only for
two of them, the flux excess starts at wavelengths shorter than 1\,mm,
but still longer than $\sim$200\,$\mu$m. In the case of the Spitzer-IRS
(5.3 -- 18\,\mic) calibration, the IRS team had to reject a
considerable fraction of candidate standards as calibrators due to IR
excess emission (Sloan 2005 \emph{, priv.\ comm.}).

\subsection{The continuous opacity source: {\rm{H$^{-}_{\rm{ff}}$}}}

\begin{figure}[!htp]
\includegraphics[height=.5\textwidth,angle=90]{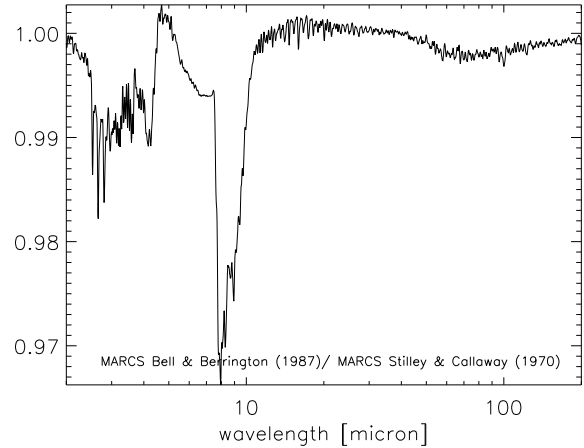}
\caption{Ratio between the synthetic spectrum calculated with the
  H$^{-}_{\rm{ff}}$ absorption coefficients of
  \citet{Bell1987JPhB...20..801B} and the spectrum based on the
  H$^{-}_{\rm{ff}}$ absorption coefficients of
  \citet{Stilley1970ApJ...160..245S}. The same stellar parameters as in
  Fig.~\ref{abooFIR} are used, data are reduced to a resolution
  of $\lambda / \Delta \lambda = 100$.}
\label{stilley}
\end{figure}

The continuous opacity of solar-type and cooler stars is dominated by
free-free absorption of the negative hydrogen ion. The {\sc marcs}
model atmosphere calculations make use of the free-free absorption
coefficients as calculated by \citet{Bell1987JPhB...20..801B}. A
quantitative assessment of the reliability of H$^{-}_{\rm{ff}}$
absorption coefficients by \citet{John1994A&A...282..890J} shows that
the absorption coefficients tabulated by
\citet{Bell1987JPhB...20..801B} are accurate to about 1\,\% for
wavelengths greater than 0.5\,$\mu$m over the temperature range
between 1400 and 10\,080\,K. \citet{Stilley1970ApJ...160..245S}
neglected the adiabatic exchange potential term yielding absorption
coefficients that may be off by $\sim$4.5\,\% at temperatures lower
than $\sim$2500\,K. Using these coefficients, one may introduce errors
in the order of 4\,\% at wavelengths $< 10$\,$\mu$m, diminishing to
less than 0.3\,\% for longer wavelengths (see Fig.~\ref{stilley}). The
prominent feature, seen around 8\,\mic\ in Fig.~\ref{stilley}, arises
from the response of SiO on the slightly different temperature structure.

\subsection{Line lists} \label{linelists}

Atomic and molecular data bases (including very different numbers of
molecular lines and based on computations spanning a wide range in
quality) are used in the construction of model
atmospheres. Knowing the limitations on the accuracy and completeness
of the used line list is of key importance when calculating
theoretical spectra. The {\sc marcs} code, with the input data used,
is particularly tuned for computations of late-type stars where
molecular opacities play an important role. As said in the
introduction of Sect.~\ref{status}, we here rely on the study done by
\citet{Decinthesis}. In this study, it has been shown that for
molecules such as CO, SiO, and OH, different molecular line lists reach a
high level of accuracy and completeness. In the case of CN, the
dissociation energy is, however, still a matter of debate \citep[see,
e.g.,][]{Lambert1994LNP...428....1L}. Most problematic is the situation
for the H$_2$O line lists. The most often used theoretical H$_2$O data bases
are the SCAN list \citep{Jorgensen2001A&A...372..249J}, the list
of \citet{PartridgeSchwenke1997}, and the very recent list of
\citet{Barber2006MNRAS.368.1087B}. These line lists are likely
to be satisfactory for opacity calculations and probably for comparing
observed and synthetic spectra at low and medium resolution, but at
higher resolution line positions based on laboratory measurements
\citep[e.g., from][]{Polyansky1997JMoSp.186..213P} should be used if
possible. 

Focusing on the wavelength range between 60 and 210\,$\mu$m (covered
by the ESA PACS instrument onboard Herschel, launch foreseen in 2008),
one should realise that the atomic
VALD{\footnote{http://ams.astro.univie.ac.at/vald/}} database
presently only contains 1380 lines with $\lambda < 122$\,$\mu$m, the
NIST
database{\footnote{http://physics.nist.gov/cgi-bin/AtData/main\_asd}}
only 90 atomic lines, while the atomic line list of
\citet{Vanhoof1998sese.conf...67V} tabulates 13\,527 lines, of which
the oscillator strengths are only known for 1660.

{\sc marcs} model atmospheres were used by, among others,
  \citet{Decin2000A&A...364..137D} and \citet{Ryde2002A&A...386..874R}
  to predict IR spectra of cool stars. In their comparison between the
  high-resolution ($\lambda/\Delta \lambda \sim 60\,000$) Fourier
  Transform Spectrometer (FTS) spectrum of $\alpha$ Boo and
  theoretical {\sc marcs} predictions, \citet{Decin2000b} show that,
  in the case of this well-known giant, the model predictions in the
  0.9 -- 5.3\,\mic\ range only differ from the observational data by
  1--2\,\%! In general, IR model spectra predictions in the wavelength
  range $< 25$\,\mic\ for A--G dwarfs and for non-pulsating K--M0
  giants may be as accurate as 3\,\% at medium resolution
  ($\lambda/\Delta \lambda \sim 1500$) and as $\sim$5\,\% at high
  resolution ($\lambda/\Delta \lambda \sim 100\,000$), as judged from
  the available observational data \citep{Decin2000b, Decin2000d}. An
  exception may be the hydrogen-line predictions due to the
  problematic computation of the self-broadening
  \citep{Barklem2000A&A...355L...5B}. At longer wavelengths, the
  accuracy and resolution of today's modern instruments remain too
  poor to constrain the model atmosphere spectra at a few percent
  level. Luckily, the depression of flux due to the line veiling in
  the FIR is estimated to be $\la 3\,\%$ at a resolution of 1500 (see
  Fig.~\ref{abooFIR}), rendering the representation of the SED of
  stellar standards by theoretical model atmosphere spectra still
  useful. In order to avoid the propagation of inaccurate predictions
  of molecular/atomic lines to the RSRF determination, different
  spectral classes should be used.  At temperatures lower than
  3500\,K, H$_2$O becomes a dominant opacity source, excluding giant
  stars with a cooler spectral type than M0~III as standard calibrators.



\subsection{Summary}

\begin{center}
\begin{table*}[htp!]
\caption{\label{overview_uncertainties_theoretical} {\bf Summary of
uncertainties attributed to the theoretical atmosphere spectra in the
2--200\,\mic\ range at a resolution $\lambda/\Delta \lambda \sim
100$.}}
\begin{tabular}{llll}\hline \hline
  \rule[-3mm]{0mm}{8mm}Description & Uncertainty & Spectral Type
  & Wavelength Region \\ 

\hline \rule[-0mm]{0mm}{5mm}$\bullet$ dependency on stellar parameters
& & & \\
\phantom{$\bullet$ }$\rightarrow$ molecular features & up to 8\,\% &
  G--K & around 2.3, 4.0, 4.2, 8\,\mic \\

\phantom{$\bullet$ }$\rightarrow$ continuum & up to 4\,\% & A--M &
  2--200\,\mic \\

$\bullet$ uncertainties on $T(\tau)$ & & & \\

\phantom{$\bullet$ } continuum flux (without high-resolution data
constraints) & $\la$3.5\,\% & A--M & 2--200\,\mic \\

\phantom{$\bullet$ } continuum flux (with high-resolution data
constraints) & 1--2\,\% & A--M & 2--200\,\mic \\

$\bullet$ presence of chromosphere/ionized wind & $\ga$10\,\% & G--M &
$\lambda > 100$\,\mic \\

\phantom{$\bullet$ }presence of circumstellar dust & $\ga$10\,\% &
A--M & $\lambda > 2$\,\mic \\

$\bullet$ continuous opacity by H$^-_{\rm{ff}}$ & 1\,\% & A--M & 2--200\,\mic \\

\rule[-3mm]{0mm}{3mm}$\bullet$ line lists & $\la$3\,\% & A0-M0 & 2--200\,\mic \\

\hline
\rule[-0mm]{0mm}{5mm}   &
1--2\,\% & A0--M0 & near-IR\\ 
\raisebox{1.5ex}[0pt]{{\sc overall budget:} for approved standards} & $\sim$3\,\% & A0--M0 & mid-IR \\
\rule[-3mm]{0mm}{3mm}\raisebox{1.5ex}[0pt]{with high-resolution data constraints}& $\sim$5\,\% & A0--M0 & far-IR\\

\hline
  \end{tabular}
\end{table*}
\end{center}

In last few sub-sections many uncertainties on the calculations of
model atmospheres are summarised  (see
Table~\ref{overview_uncertainties_theoretical} for an overview), and
the total uncertainty is not (as often done) estimated as the
root-sum-square of all above-mentioned uncertainties since they are
often mutually dependent.  The total uncertainty on the stellar
reference SEDs depends on the spectral type of the target, the
wavelength range under study and the instrumental resolution.  In case
the standard star has been properly studied using high-resolution
optical and/or near-IR data and the presence of an IR excess can be
excluded, one can constrain \Teff\ and the temperature distribution to
within $\sim$50\,K, yielding near-IR (continuum + line)
medium-resolution predictions better than 1--2\,\%, mid-IR predictions
better than $\sim$3\,\% and far-IR predictions better than $\sim$5\,\%
for stars with earlier spectral type than M0. Each of the discussed
spectral classes (early A dwarfs, solar analogs, and G9-M0 giants) has
its own drawbacks. Consequently, with the aim of constraining the RSRF
at a few percent level, the selected stellar standard candles should
cover a wide diversity in spectral types to avoid biases to spectral
features typical of different spectral classes.

\section{Comparison between different IR stellar reference SEDs}
\label{comparison} 

\begin{figure*}[!thp]
  \sidecaption
  \includegraphics[width=12cm,angle=0]{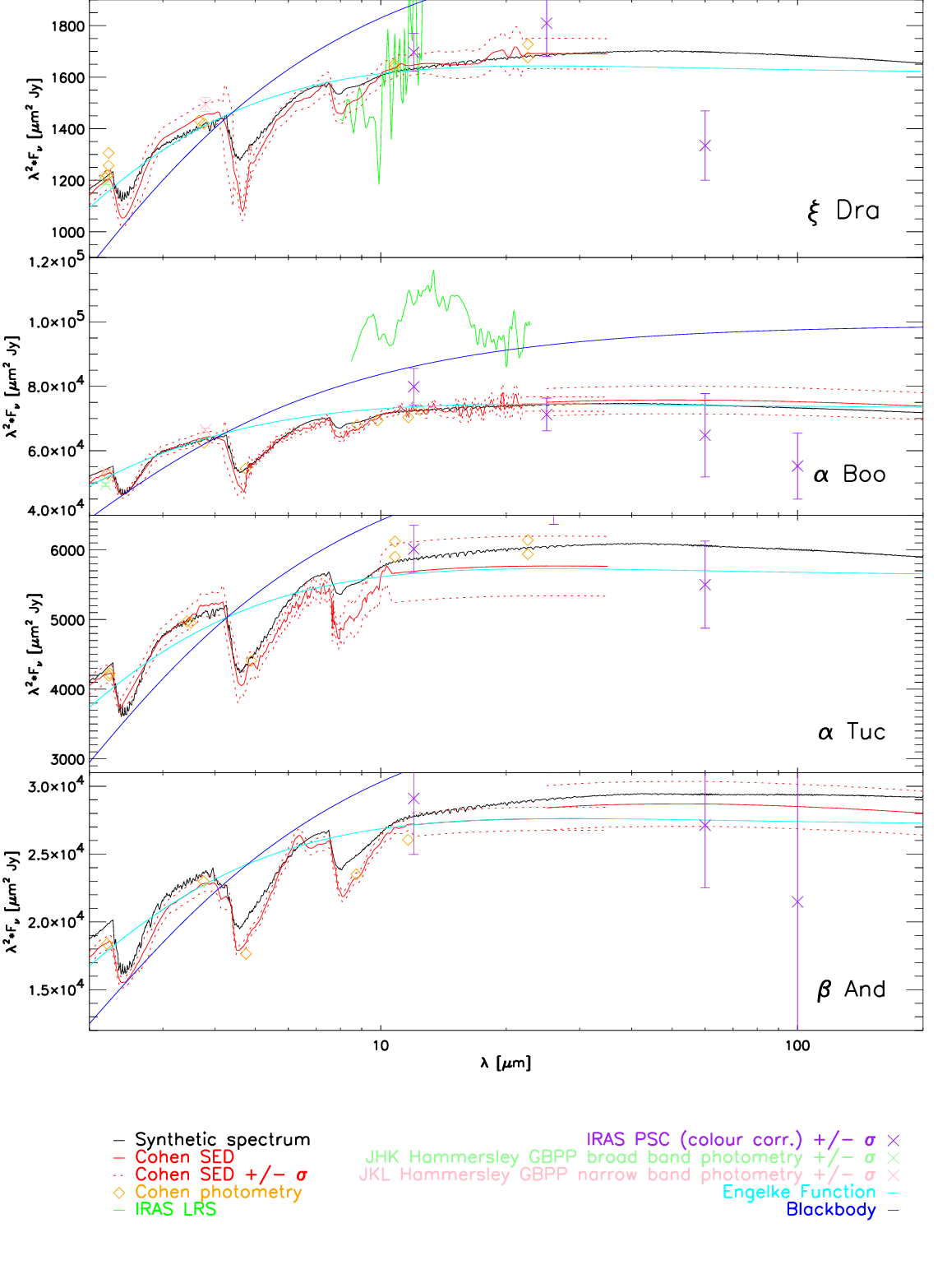}
  \caption{Comparison between the various reference SEDs discussed in
  this paper in the 2 to 200\,$\mu$m wavelength
  range. \object{$\alpha$ Boo} (K2~IIIp) and \object{$\beta$ And}
  (M0~III) represent two composites from the CWW network, and
  \object{$\xi$ Dra} (K2~III) and \object{$\alpha$ Tuc} (K3~III)
  represent two templates. {\sc marcs} model atmosphere spectra are
  displayed at a resolution $\lambda / \Delta \lambda = 100$. IRAS LRS
  and PSC data and several photometric data points collected from the
  ISO GBPP are also shown.}
  \label{figcompare} 
\end{figure*}

It is instructive to compare the different IR stellar reference SEDs
presented in this paper. We therefore have chosen to compare the
proper blackbody, Engelke function, and model atmosphere spectrum with
the spectra of two \emph{composites} (as explained in
Sect.~\ref{Cohen}) (\object{$\alpha$ Boo} (K2~IIIp) and
\object{$\beta$ And} (M0~III)) and two \emph{templates} (\object{$\xi$
Dra} (K2~III) and \object{$\alpha$ Tuc} (K3~III)) from the sample of
CWW (see Fig.~\ref{figcompare}). Figure~\ref{figcompare} also plots the
IRAS LRS and PSC data and different photometric data points collected
from the ISO Ground Based Preparatory Programme
(GBPP){\footnote{http://www.iso.vilspa.esa.es/users/expl\_lib/ISO/wwwcal/\\isoprep/gbpp/photom/}}
as obtained by \citet{Hammersley1998A&AS..128..207H} and
\citet{Hammersley2003clim.conf..129H}.

For the {\sc marcs} atmosphere spectra, stellar parameters as
determined by \citet{Decin2000d} were used as input parameters. Main
parameters for the absolute flux, \Teff, and \ad, are listed in
Table~\ref{list_teff_and_ad}. For the computation of the flux values of the
Engelke function for both {\em composites}, \Teff\ and \ad, as given
in the headers by CWW, are used. Note that for \object{$\beta$ And},
the re-scaled (by CWW) angular diameter of 13.71\,mas is used: based
on the InfraRed Flux Method (IRFM),
\citet{Blackwell1991A&A...245..567B} determined a lower angular
diameter of 13.219\,mas corresponding to the stellar temperature of
3839\,K. As noted in Sect.~\ref{Engelke}, the Engelke function needs a
higher angular diameter (or \Teff) for low-gravity, low-temperature
giants in order to attain the correct flux level. For both {\em
templates}, same values as for the {\sc marcs} spectrum are used. The
input for the blackbody calculation is the same as for the Engelke
function.

\begin{center}
\begin{table}[htp!]
  \caption{\label{list_teff_and_ad} Stellar parameters used to for the
calculation of {\em (1.)} the {\sc marcs} model atmosphere spectra
(columns 2\,--\,5) and {\em (2.)}  the Engelke function and blackbody
(BB) (columns 6\,--\,7) in Fig.~\ref{figcompare}. Effective
temperature, \Teff, is given in Kelvin, gravity, $g$, in cm/s$^2$, and
angular diameter, \ad, in mas.}
    \begin{tabular}{lcccccc}\hline \hline
      \rule[-0mm]{0mm}{5mm} & \multicolumn{4}{c}{\sc marcs} &
      \multicolumn{2}{c}{Engelke and BB} \\
      \raisebox{1.5ex}[0pt]{target}& \Teff & $\log g$ & [Fe/H] & \ad & \Teff & \ad \\
      \hline
      \rule[-0mm]{0mm}{5mm}\object{$\xi$ Dra} & 4440 & 2.40 & 0.10 & 3.09 &  4440 & 3.09 \\
      \object{$\alpha$ Boo} & 4320 & 1.50 & $-0.50$ & 20.66 & 4362 & 21.12 \\
      \object{$\alpha$ Tuc} & 4300 & 1.35 & 0.00 & 5.90 & 4300 & 5.90 \\ 
      \rule[-3mm]{0mm}{3mm}\object{$\beta$ And} & 3880 & 0.95 & 0.00 & 13.48 & 3839 & 13.71 \\
      \hline
  \end{tabular}
\end{table}
\end{center}

\begin{figure}[!htp]
\includegraphics[height=.5\textwidth,angle=90]{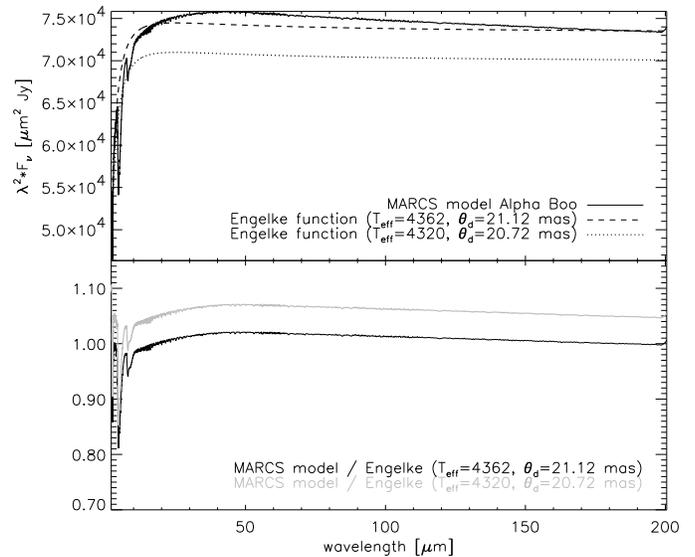}
\vspace*{.5ex}
\caption{{\em Upper panel:} Comparison between the {\sc marcs} model
spectrum of \object{$\alpha$ Boo} ($\lambda / \Delta \lambda = 100$)
and predictions using the Engelke function with parameters
\Teff\,=\,4362\,K and \ad\,=\,21.12\,mas (dashed line), and
\Teff\,=\,4320\,K and \ad\,=\,20.72\,mas (dotted line). {\em Bottom
panel:} Ratio between {\sc marcs} model predictions and both Engelke
functions.}
\label{marcsengelke}
\end{figure}

Concentrating on the two composites, \object{$\alpha$ Boo} and
\object{$\beta$ And}, of which a large part of the CWW spectrum
consists of observational data, it is immediately clear that the
Engelke function does a much better job than the blackbody function of
representing the observed continuum stellar SED.  The simple blackbody
function can not even be used to estimate broad-band features in the
RSRF. Between $\sim$23 and $\sim35$\,$\mu$m, the CWW spectrum of both
composites is given by their Engelke function. As discussed in
Figs.~\ref{Engelke_PP} and \ref{PPversusSPH}, the shape of the
continuum flux differs between a proper {\sc marcs} model spectrum and
the flux values given by the Engelke function, the reason being the
neglect of the influence of the gravity and sphericity effects, which
are important for giant stellar atmospheres. Between 20 and
200\,$\mu$m, the maximum difference between the Engelke function used
by CWW and the {\sc marcs} model atmosphere for \object{$\alpha$ Boo}
is near 60\,$\mu$m, where it attains $\sim$2.5\,\%, rising to
$\sim$7.5\,\% when the same \Teff\ and \ad\ are used as for the
{\sc marcs} model spectrum (Fig.~\ref{marcsengelke}).

The large absolute difference between the IRAS LRS spectrum of
$\alpha$ Boo and both the {\sc marcs} model atmosphere spectrum and
CWW composite \citep[$\sim 35\,\%$, see
also][]{VanMalderen2004A&A...414..677V} remains unexplained.  The LRS
raw data are extracted from the Groningen IRAS database and calibrated
with the {\sc lrscal} routine in the {\sc gipsy} package.  Although
the IRAS LRS data were originally not meant to be absoluted
calibrated, absolute calibration factors were determined by
\citet{Volk1989AJ.....98.1918V} and \citet{Cohen1992AJ....104.2030C}.
The same procedure was used by \citet{Cohen1996AJ....112.2274C}, who
however only needed a factor of 0.95 to splice the LRS data to the
CGS3 data between 7.5 and 13\,\mic.

\begin{figure}[!htp]
\includegraphics[height=.5\textwidth,angle=90]{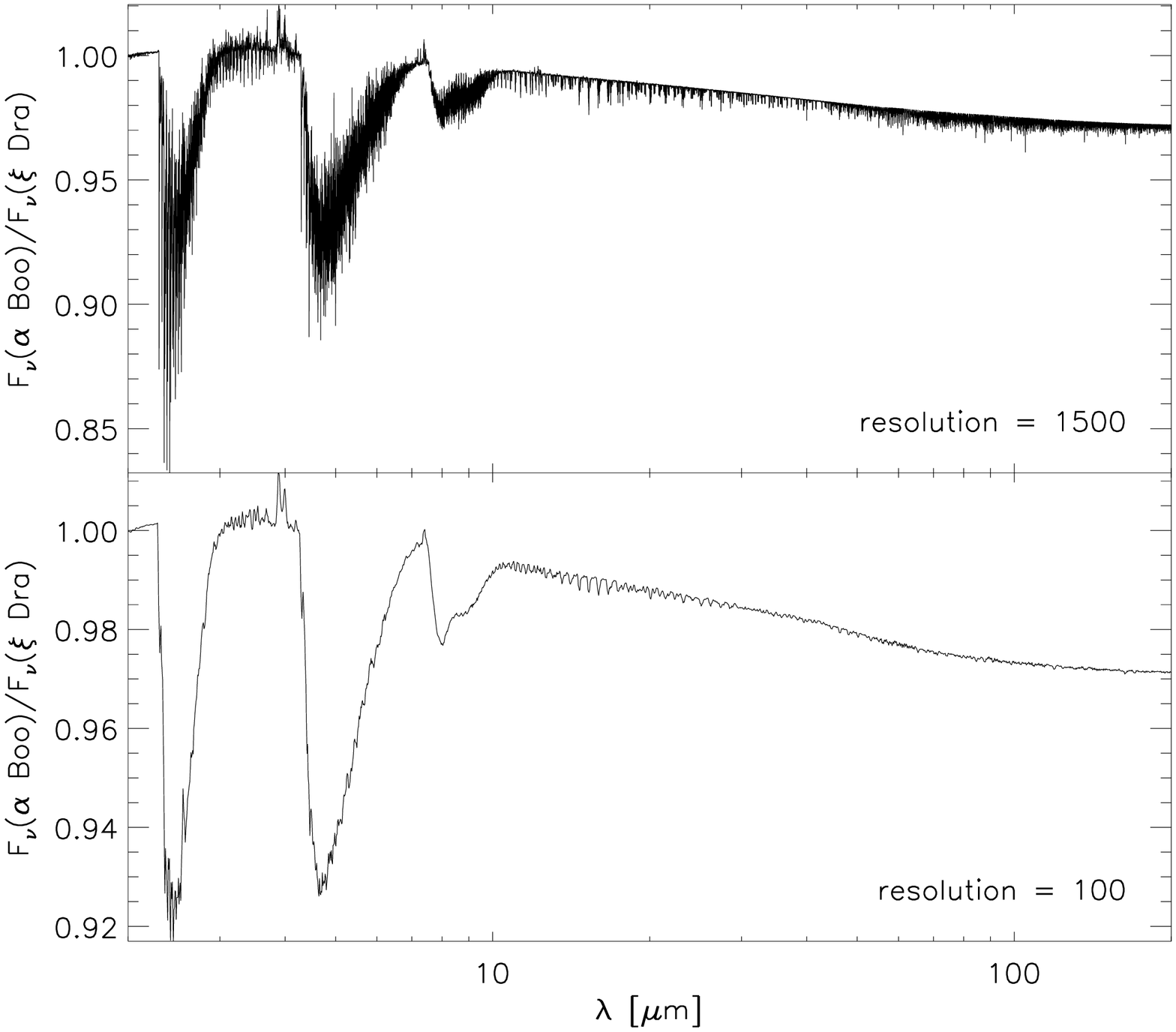}
\vspace*{.5ex}
\caption{ {\em Upper panel:} Comparison between the {\sc marcs} model
spectrum of \object{$\alpha$ Boo} and \object{$\xi$ Dra} at a
resolution of 1500. Both spectra have a fictitious angular diameter of
1\,mas. {\em Bottom panel:} Same as for the upper panel, but
now at a resolution of 100. In regions of strong molecular absorption,
the maximum difference is 15\,\% for $\lambda/\Delta \lambda =
1500$, and 8\,\% for $\lambda/\Delta \lambda = 100$.}
\label{ksidra_versus_aboo}
\end{figure}

Inspecting the molecular absorption features in both templates,
\object{$\xi$ Dra} and \object{$\alpha$ Tuc}, a clear difference is
visible between the template spectrum (red) and the {\sc marcs} model
atmospheres (black). As discussed in \citet{Decin2000A&A...364..137D}
and in Sect~\ref{dependency}, differences in abundance pattern (here
mainly the C and O abundance), temperature, gravity, and metallicity
result in substantial differences in strength in molecular
absorption. In the case of \object{$\xi$ Dra}, the composite of
\object{$\alpha$ Boo} was used to construct the template, the
composite of \object{$\alpha$ Hya} (K2~II-III) was used for
\object{$\alpha$ Tuc}. Inspecting the literature study of
\object{$\alpha$ Boo} and $\xi$ Dra presented in Appendix~D in
\citet{Decin2000d} clearly shows that independent studies
incorporating both objects in general yield an effective temperature
higher by $\sim$120\,K, a logarithm of the gravity higher by
$\sim$1\,dex, and a metallicity [Fe/H] higher by $\sim$0.50\,dex for
$\xi$ Dra with respect to \object{$\alpha$ Boo}.
Figure~\ref{ksidra_versus_aboo} compares the {\sc marcs} flux
predictions of $\alpha$ Boo and $\xi$ Dra. Especially in the regions
of molecular absorption, the (relative) difference between both
spectra may be 10\,\% or even larger, explaining why the difference
between the CWW spectrum and the {\sc marcs} model spectrum is larger
than the uncertainties quoted by CWW.  Consequently, the use of the
composite spectrum of the metal-deficient K2-gaint $\alpha$ Boo as
a template to represent a whole class of K2 giants in the CWW network
will introduce additional uncertainties, which may propagate through
the RSRF determination.

For both composites, the far-IR extrapolations constructed by
\citet{Cohen1996AJ....112.2274C} differ by a few percent from
the {\sc marcs} atmospheric predictions, since the {\sc marcs} models
are computed in spherical geometry, while the CWW predictions are
interpolations in a grid of plane-parallel models. For the cooler,
more-extended giant calibrators, sphericity effects may yield an infrared flux
excess in the order of $\sim$5\,\% (see Fig.~\ref{PPversusSPH}).

\section{Theoretical atmosphere spectra for stellar candles and how to
  obtain them} 

 The four reference SEDs constructed from the {\sc marcs} atmosphere
code presented in Fig.~\ref{figcompare} are available via the
electronic version of the article. Other reference SEDs constructed in
the framework of the calibration of e.g. ISO, Spitzer, MIRI etc.\ are
available upon request. If one needs extra reference SEDs for
additional stellar calibration sources, one may always contact the
authors, who are willing to provide you with the appropriate
theoretical model atmosphere spectrum in the wavelength range and at
the resolution requested. 

\section{Conclusions} \label{conclusions}

In this paper, various sets of reference SEDs used for 
determining  the spectrophotometric calibration of IR
spectrometers (onboard satellites) have been discussed. Our main emphasis
was on the {\em stellar} reference SEDs, with special focus on model
atmosphere spectra.  It is shown that the predicted medium-resolution
IR model atmosphere spectra are accurate within 1--2\,\% in the
near-IR, $\sim$3\,\% in the mid-IR, and $\sim$5\,\% in the far-IR for
stars with spectral types earlier than M0.  From the four types of
stellar reference SEDs discussed in this study (blackbody, Engelke
function, templates in the CWW network, and theoretical atmosphere
spectra), it is believed that theoretical atmosphere spectra make up
the best representations nowadays for the stellar SED, especially in
case one wants to calibrate instruments with a spectral resolution
$\ga 500$. We note, however, that at $\lambda > 20$\,\,\mic, stellar
calibrators might be too faint to be used in the spectrophotometric
calibration pedigree.

Since the ultimate goal of the calibration system is often to be
capable of deriving spectral flux values that are trustworthy to 3\,\%
or better on both absolute and relative scales, one should aim at
building a highly accurate system of stellar reference SEDs. A good
sky coverage by the calibrators is an important ingredient in terms of
a time-efficient determination of the (spectrophotometric)
calibration. In that sense, the IR network constructed by CWW is a
good starting point. However, the CWW network has its limitations in
terms of accuracy at representing the (molecular) spectral features
and the SED at wavelengths longer than $\sim$20\,$\mu$m. For a list of
candidate calibrators, one therefore should put effort into obtaining
ancillary observational data to both (1) constrain the reliability of
the candidates as standard stellar sources and (2) estimate the
stellar parameters to compute a set of highly reliable theoretical
model atmosphere SEDs. This set should compromise standard stellar
candles with different spectral types ranging between A0~V and
M0~III. Experiences with the Short-Wavelength Spectrometer (SWS)
onboard ISO and the InfraRed Spectrometer (IRS) onboard Spitzer have
shown that this set should consist of some ten to fifteen stellar
calibrators.  In the framework of the calibration plans for PACS
onboard the ESA-Herschel satellite and MIRI onboard the NASA-JWST
satellite, a set compromising theoretical spectra of $\sim$15 stellar
calibrators is in preparation \citep{DecinBauwens2006}. The input
stellar parameters for each set of model spectra have to be estimated
from a proper analysis of high-resolution optical or near-IR data. A
systematic cross-calibration with planets and asteroids is the last
step in the spectrophotometric calibration pedigree. Only then can a
highly accurate calibration system be developed.



\begin{acknowledgements}LD acknowledges financial support from the
  Fund for Scientific Research - Flanders (Belgium) and KE from the
 Swedish Research Council.  We are grateful to B.\ Gustafsson, B.\
 Edvardsson, and B.\ Plez for their ongoing support when using the
 {\sc marcs} model atmosphere code developed at the Uppsala
 University. LD thanks C.\ Waelkens, B.\ Vandenbussche, T.\ Verhoelst,
 J.\ Blommaert, and colleagues from the ISO, Spitzer, Herschel, and
 MIRI calibration teams for many fruitful discussions on the use of
 the most appropriate SEDs to represent the spectrum of standard stars
 used in the spectrophotometric calibration process of these infrared
 instruments.  T.\ Verhoelst is thanked for Fig.~\ref{FigTijl}.
\end{acknowledgements} 

%

\end{document}